\documentclass[10pt]{article}

\usepackage{amsmath}
\usepackage{amssymb}

\usepackage{graphicx}

\usepackage{cite}

\usepackage{color}

\usepackage[labelfont=bf,labelsep=period,justification=raggedright]{caption}

\usepackage{subfig}
\usepackage[percent]{overpic}

\makeatletter
\renewcommand{\@biblabel}[1]{\quad#1.}
\makeatother

\date{}

\begin{document}

\begin{flushleft}
{\Large \textbf{The Decentralized Structure of Collective Attention on the Web}}
\\
Lingfei Wu$^{1}$,
Jiang Zhang$^{2}$$\ast$,
\\
\bf{1} Department of Media and Communication, City University of Hong Kong, Hong Kong, China
\\
\bf{2} Department of Systems Science, School of Management, Beijing Normal University, Beijing, China
\\
$\ast$ E-mail: zhangjiang@bnu.edu.cn
\end{flushleft}

\section*{Abstract}

\textit{Background}: The collective browsing behavior of users gives rise to a flow network transporting attention between websites. By analyzing the structure of this network we uncovered a non-trivial scaling regularity concerning the impact of websites.

\textit{Methodology}: We constructed three clickstreams networks, whose nodes were websites and edges were formed by the users¡¦ switching between sites. We developed an indicator $C_{i}$ as a measure of the impact of site $i$ and investigated its correlation with the traffic of the site ($A_{i}$) both on the three networks and across the language communities within the networks.

\textit{Conclusions}: We found that the impact of websites increased slower than their traffic. Specifically, there existed a relationship ${C_{i}}\sim{A_{i}}^{\gamma} (\gamma < 1)$. We suggested that this scaling relationship characterized the decentralized structure of the clickstream circulation: the World Wide Web is a system that favors small sites in reassigning the collective attention of users.



\section*{Introduction}

The explosive growth of the World Wide Web in the past two decades presents an urgent challenge for developing a quantitative, predictive theory of the interaction between the Web and users. Previous studies analyzed the structure of the hyperlink network \cite{watts1998collective, broder2000graph, kleinberg2001structure,kleinberg2000navigation,page1999pagerank} and also the individual browsing records \cite{meiss2010agents,qiu2005analysis,chmiel2009scaling,white2010assessing} in order to investigate the collective surfing behavior. While these studies paved way for the following research, they have limitations restricted by the data. Firstly, as pointed out in \cite{meiss2008ranking}, hyperlinks are too simple to represent the rich interactions between sites as a result of users' various online activities. From bookmarks and default home pages to historical viewing records, there are many different ways in which clickstreams are generated between sites of no hyperlink connections \cite{meiss2008ranking}. Secondly, although individual surfing records has been extensively investigated \cite{huberman1998strong,qiu2005analysis,chmiel2009scaling,white2010assessing}, there is still a lack of research studying collective browsing behavior from a network perspective \cite{bollen2009clickstream}.

There are generally two different opinions concerning the surfing dynamics. One is the ``rich-get-richer" paradigm, which suggests that user navigation strengthens the inequality of traffic among sites \cite{barabasi1999emergence,cho2004impact,introna2000defining}. The other is the ``egalitarian" paradigm arguing that the surfing activities of users actually makes the Web a level-play place where new sites have a greater chance of acquiring popularity \cite{fortunato2006topical}. In the current study we investigated clickstreams formed by a large number of users to examine these two effects. We collected data from Google (www.google.com) and Alexa (www.alexa.com) and constructed three website-level clickstream networks at different time points \cite{brainerd2001case,bollen2009clickstream,fortunato2006topical}. In each of the networks, the nodes were websites and the edges showed the daily percentage of global users who visited two websites successively. We defined $A_{i}$ as the traffic of site $i$ and $C_{i}$ the impact of the site on the rest of sites in clickstream circulation. If $C_{i}$ increases faster than $A_{i}$, the ``rich-get-richer" effect is supported; otherwise, the ``egalitarian" effect is supported. It turned out that $C_{i}$ scales sublinearly with $A_{i}$ as ${C_{i}}\sim{A_{i}}^{\gamma} (\gamma < 1)$. This scaling pattern was observed to be universal, existing both in the three clickstream networks and across the language communities within the networks. We suggested that this pattern, as an evidence of the ``egalitarian" effect, resulted from the decentralized structure of clickstream networks. That is, compared to large sites, small sites had a disproportionately larger impact in the circulation of clickstreams.

We would like to point out that the presented approach of clickstream network analysis is not only interesting at its own right, but also provides a new method investigating various online activities. For example, traditional studies on news diffusion focused on the diffusion of news among users \cite{funkhouser1971rise,lerman2010information}, but from the perspective of this approach, we can also understand the diffusion process in a ``reversed" way, that is, the allocation and transmission of users' attention among news \cite{wu2007novelty}. Therefore, the rise and decay of news is the result of the competition among them for users' collective attention \cite{wu2007novelty}. Obviously, News can also be tags \cite{cattuto2009collective}, videos \cite{wu2009feedback} or any other type of information resources when clickstream network is applied to analyze a specific type of online activity.


\section*{Materials and Methods}

\subsection*{Data collection}

We at first selected three lists of top 1000 sites at different time points. Two of them were selected from Google statistics (http://www.google.com/adplanner/static/top1000/) and the rest one was selected from Alexa reports (http://www.alexa.com)(please refer to Supplemental Materials for the detailed information of these lists). We then downloaded from Alexa the clickstreams between the sites on the lists. From the downloaded data we constructed three clickstream networks (which are called $w1$, $w2$, and $w3$ hereafter), in which a directed, weighted edge from nodes $i$ to $j$ indicated the daily percentage of the global Web users who visited $i$ and $j$ successively. It should be noted that as Alexa only reports a maximum of ten top inbound and outbound clickstreams for each site, our dataset does not neccessarily include all the clickstreams between the studied sites. We actually constructed and studied the ``backbone networks" of the clickstreams on the Web \cite{foti2011nonparametric}. That is, we extracted the top clickstreams connecting the largest sites on the Web. Essential statistics of the three networks, including degree distribution, distribution of weights, and weighted degree distribution, are shown in the Supplementary Materials. In Fig.\ref{fig.1} we plot $w2$ as an example of the clickstream networks, in which we also show the language-based communities of sites to be introduced in the section of Results.

\begin{figure}[!ht]
\begin{center}
\includegraphics[scale=0.6]{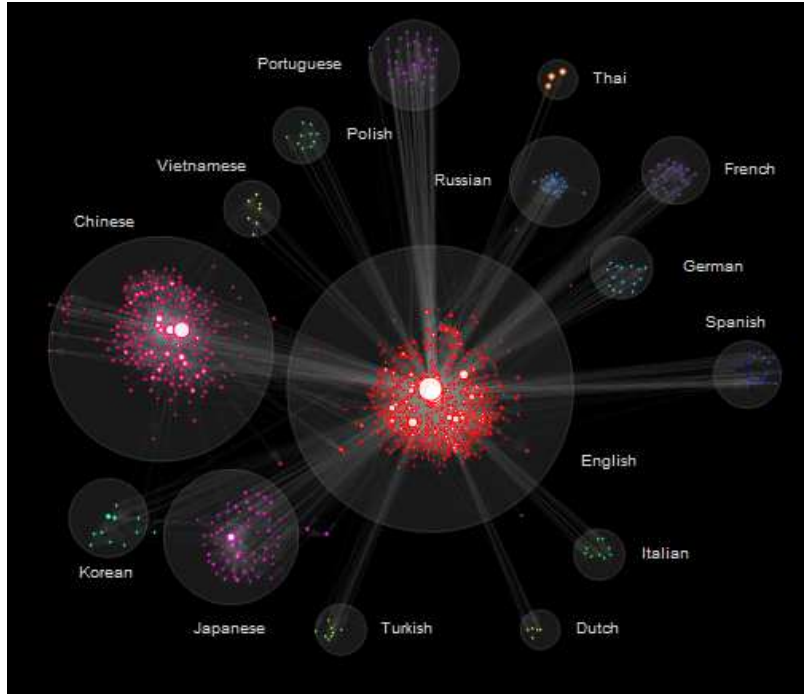}
\end{center}
\caption{{\bf The visualization of $w2$.} Small white circles represent websites, big gray circles correspond to language communities, and edges show the clickstreams. Websites of the same language are placed together within a big gray circle and are assigned the same color of edges, with their traffic being reflected by the size of small circles. In calculating the coordinates of the websites, we designed a new algorithm called ``two-level spring embedding algorithm", which used spring embedder for two times. Firstly we aggregated $w2$ into a community-level network, in which nodes were communities and edges were the clickstreams transported between them. We visualized this network using a spring embedding algorithm \cite{fruchterman1991graph} but only plotted the nodes, which are exactly the gray circles in the figure, whose size is proportional to the total amount of clickstream within a community. Secondly we applied the said spring embedding algorithm algorithm on each of the communities and rescaled the coordinates of websites in order to place them in the gray circles. We found that with the help of the ``two-level spring embedding algorithm" we could show the communities particulary clear while remaining the topological structure of the entire network.}
\label{fig.1}
\end{figure}

\subsection*{The definition of $A_{i}$, $C_{i}$, and $\gamma$}

\begin{center}
  \begin{figure*}[!ht]
  \centering
      \subfloat[\label{subfig-1:a}]{%
      \begin{overpic}[scale=0.7]{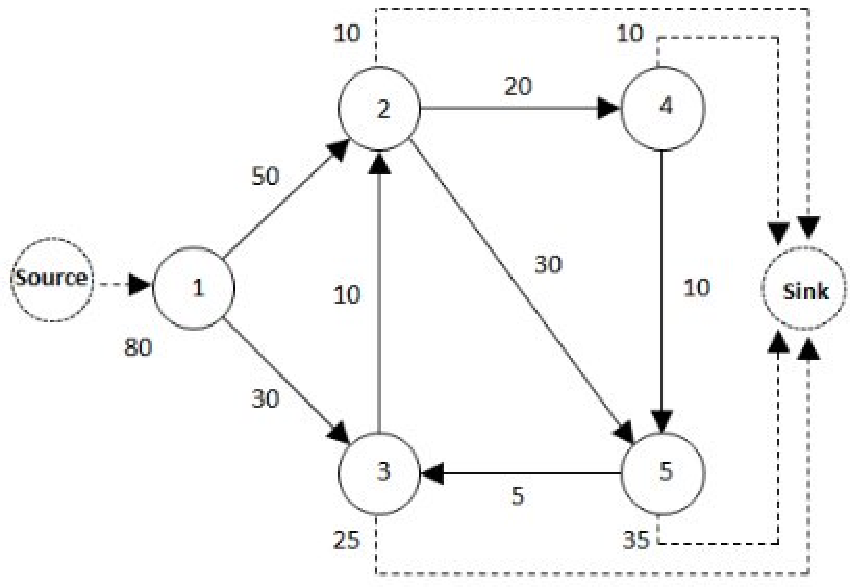}
      \end{overpic}
    }
       \subfloat[\label{subfig-2:b}]{%
      \begin{overpic}[scale=0.6]{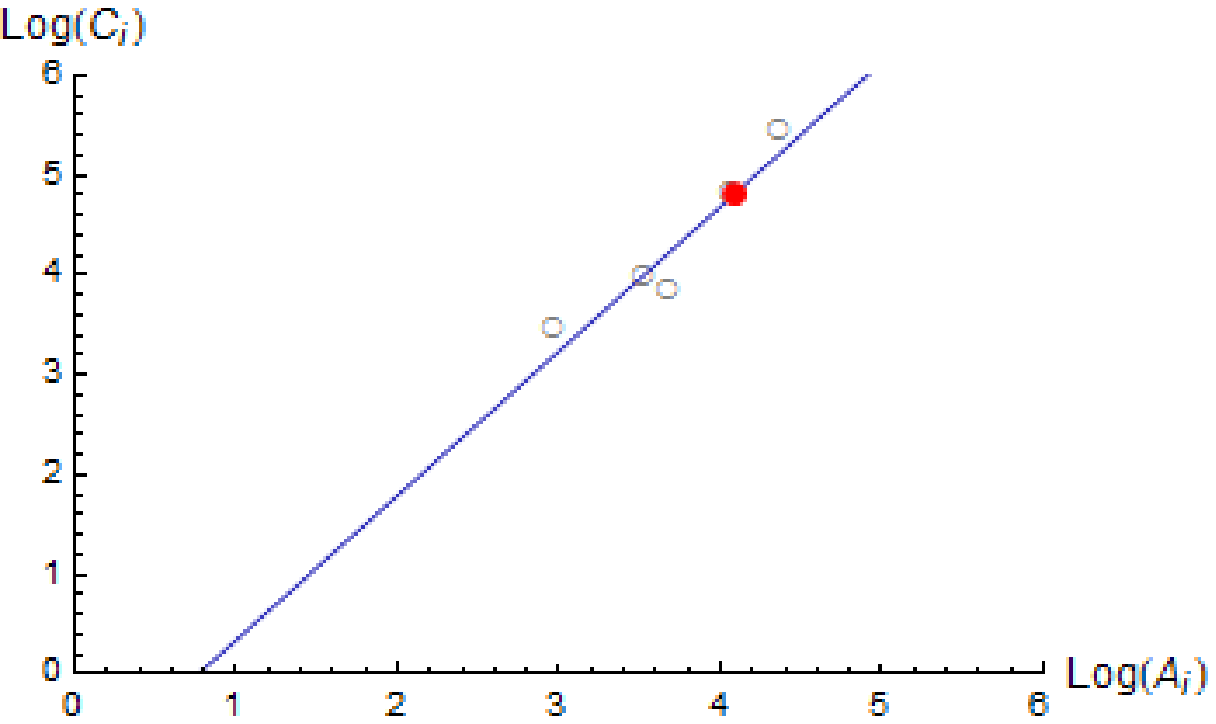}
      \end{overpic}
    }
\caption{{\bf An example clickstream network (a) and the fittings of $\gamma$ in the example network (b).} The red node denotes the values of $A_{2}$ and $C_{2}$.}
    \label{fig.2}
  \end{figure*}
 \end{center}

\begin{figure}[!ht]
\begin{center}
\includegraphics[scale=0.6]{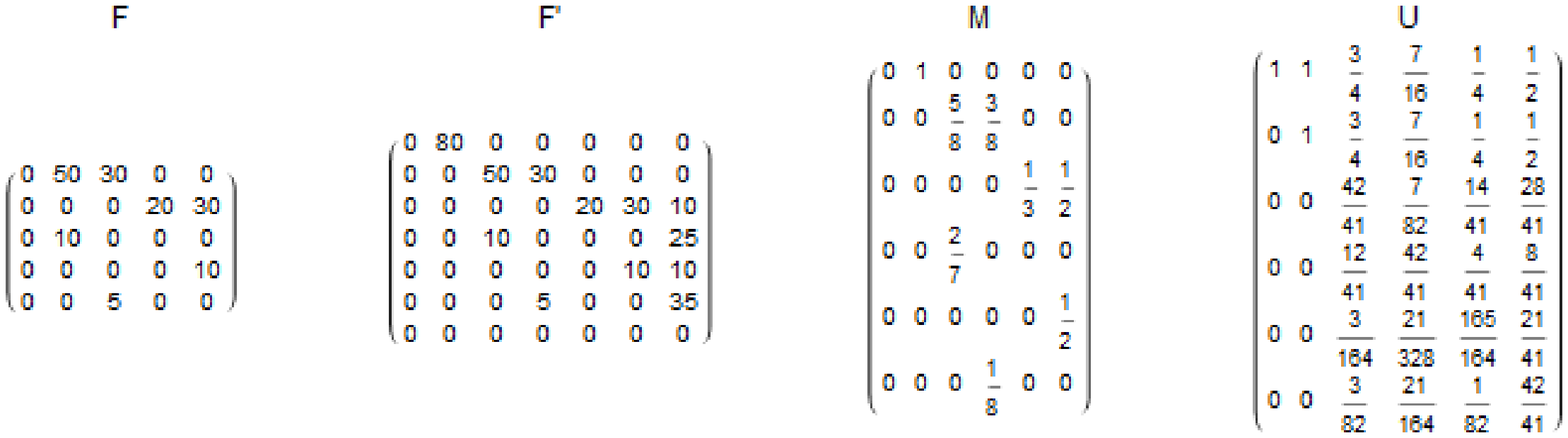}
\end{center}
\caption{{\bf Summary of the steps in deriving matrix $U$.}}
\begin{flushleft}
\end{flushleft}
\label{fig.3}
\end{figure}

To show how to calculate $A_{i}$, $C_{i}$, and $\gamma$ exactly, let us take a look at an example clickstream network provided by Fig.\ref{fig.2}a. In this network nodes are websites, edges are clickstreams, and the weights of edges reflect the number of distinct users that hop between sites. We balance the network (dashed lines) by adding two artificial nodes, ``source" and ``sink", to make sure that at each node the sum of inbound and outbound streams are equal \cite{higashi1986extended}. Then we normalize the matrix form of the balanced network $F'$ by row to obtain the transition matrix $M$ as given by Eq.\ref{eq.1}. An element $m_{ij}$ of $M$ denotes the probability that a random user visits website $i$ and $j$ successively. Note that there are only $n+1$ rows (columns) in $M$
for the row (column) corresponding to ``sink'' should be removed in order to derive Eq.\ref{eq.5}.
\begin{equation}
\label{eq.1}
m_{ij}=\frac{f'_{ij}}{\sum_{k=1}^{n+1}f'_{ik}},\hspace{1cm}\forall
i,j=0,1,\cdot \cdot \cdot,n
\end{equation}
Now we define $A_{i}$ and $C_{i}$ as follows
\begin{equation}
\label{eq.2} A_{i}=\sum_{k=1}^{n+1}f'_{ik},\hspace{1cm}\forall
i=1,2,\cdot \cdot \cdot n
\end{equation}
\begin{equation}
\label{eq.3} C_{i}=G_{i}\sum_{k=1}^{n}u_{ik},\hspace{1cm}\forall
i=1,2,\cdot\cdot\cdot n
\end{equation}
In Eq.\ref{eq.3}, $u_{ij}$ is the element of
\begin{equation}
\label{eq.4} U = \frac{1} {I-M} = I + M + M^{2}+ \dots +M^{\infty}
\end{equation}
and $G_i$ is defined as:
\begin{equation}
\label{eq.5} G_{i}=\frac{\sum_{j=1}^nf'_{0j}u_{ji}} {u_{ii}}.
\end{equation}
Where $f'_{0j}$ is the balanced flow from ``source" to $j$. From
Eq.\ref{eq.4} and Eq.\ref{eq.5} we know that $u_{ij}$ calculates the
fraction of the total flow from $i$ to $j$ along all possible pathes
over the total traffic of $i$, and thus $G_i$ is total flow
transported from ``source" to $i$ \cite{barber1978markovian} summed
over all possible pathes (excluding the flow on the self-loop of
$i$). Using the data of the example network, Figure.\ref{fig.3}
gives a summary of the above mentioned calculations, which prepares
data for the testing of the scaling relationship:

\begin{equation}
\label{scaling} {C_{i}}\sim{A_{i}}^{\gamma}.
\end{equation}

In sum, $A_{i}$ stands for the traffic of an arbitrary site $i$ in
the balanced clickstream network. $C_{i}$ reflects the circulated
clickstreams moderated by $i$, in specifically, the total number of
users who have visited this site and still remain in the network.
Therefore, we can treat $C_{i}$ as a measure of the total (both
direct and indirect) impact of site $i$ on the rest of sites in
clickstream circulation \cite{higashi1993network,zhang2010scaling}.
$\gamma$ measures the increase of impact with traffic, averaged over
all sites. We can interpret $\gamma$ as the level at which large
sites dominate the circulation of clickstreams
\cite{vitali2011network}. For example, suppose we have two
clickstream networks of the same traffic distribution but are
different in $\gamma$, $A_{i}$ = \{1, 2, 3, 4, 5\}, $\gamma' = 1/2$,
and $\gamma'' = 2$. We can derive that $C_{i}'$ = \{1, 1.4, 1.7, 2,
2.2\} and $C_{i}''$ = \{1, 4, 9, 16, 25\}. The impact of the largest
node is $2.2$ in the former network, meaning that it controls
$(2.2/(1+1.4+1.7+2.2))\approx 27\%$ circulated clickstreams.
Similarly, we can also derive that in the latter network the largest
node controls $45\%$ circulated clickstreams, leading us to the
conclusion that the latter network is more centralized. To conclude,
$\gamma < 1$ implies a ``democratic" flow structure in which the
impact of websites are evenly distributed, whereas $\gamma > 1$ is
the signature of a ``oligarchic" flow structure in which a small
group of ``hubs" controls the entire network.

As suggested in \cite{garlaschelli2003universal,warton2006bivariate,zhang2010scaling}, a scaling relationship between $A_{i}$ and $C_{i}$, to be existed, allows one to regress $Log(C_i)$ on $Log(A_i)$ and obtain $\gamma$ as the slope of ordinary least square (OLS) regression. For example, Figure.\ref{fig.1}b plots $log(C_{i})$ against $log (A_{i})$ in the example network, in which the data point corresponding to node $2$ ($A_{2} = 60$; $C_{2} = 125$) is colored in red.


\section*{Results}

\subsection*{A scaling pattern that reveals the decentralized structure of clickstream networks}

\begin{figure}[!ht]
\begin{center}
\includegraphics[scale=0.7]{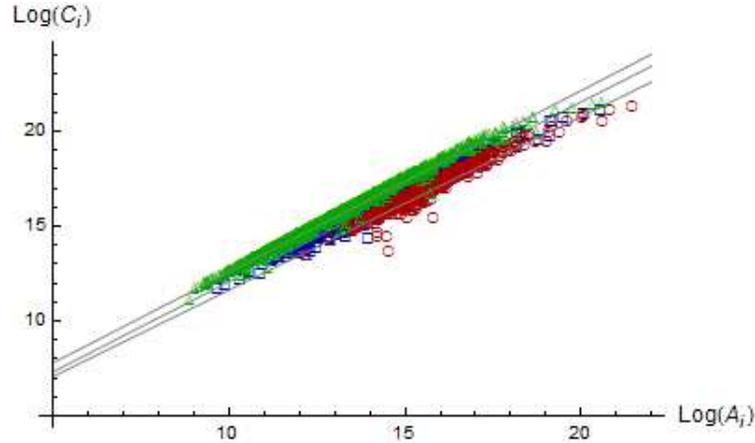}
\end{center}
\caption{{\bf The scaling relationship between $A_{i}$ and $C_{i}$ in the three clickstream networks.} The data points from three networks are plotted in different colors and styles: blue squares for $w1$, red circles for $w2$, and green triangles for $w3$. The values of $\gamma$ (black line) were $0.95$, $0.92$, and $0.96$, respectively. Please refer to Table.\ref{tab.1} for more information concerning the fitting of the three networks.}
\label{fig.4}
\end{figure}

The traffic $A_{i}$ of a site is proportional to the probability that a random user chooses it as the entrance of the virtual world \cite{page1999pagerank}, thus is widely used to indict the website popularity \cite{meiss2010agents,qiu2005analysis}. However, if we want to study the long-distance, complex interactions between sites, the investigation on the distribution of traffic \cite{qiu2005analysis,chmiel2009scaling,white2010assessing} is not enough. We should probe into the transportation of traffic between sites, that is, the flow of clickstreams \cite{brainerd2001case,bollen2009clickstream}. In the current study, to examine the ``rich-get-richer" effect against the ``egalitarian" effect, we defined the impact $C_{i}$ of a arbitrary site as the amount of the circulated clickstreams controlled by the site and investigated its relationship with the traffic $A_{i}$ on three clickstream networks (please refer to the section of Data and Method for the calculation of $A_{i}$ and $C_{i}$).

As shown by Figure.\ref{fig.4}, we found a scaling relationship ${C_{i}}\sim{A_{i}}^{\gamma}$, in which $\gamma$ was estimated to be in the range of $0.92 \sim 0.96$ (Table.\ref{tab.1}). This finding suggested that the impact of websites increases slower than its traffic, which is an evidence of the decentralized structure of the clickstream networks \cite{garlaschelli2003universal,zhang2010scaling}.

\begin{table}[!ht]
\begin{center}
\caption{The statistics of three studied clickstream networks. }
\begin{tabular}{|c|c|c|c|c|c|c|c|c|}
\hline
Network &     $N_{sites}$ & $N_{edges}$  & Daily clickstreams &  $\gamma$& $R^{2}$ of $\gamma$   \\\hline
\textit{w1}&    979&           11906&       $5.45 \times 10^{9}$&   0.95&      0.98 \\
\textit{w2}&    956&           11529&       $1.38 \times 10^{10}$&   0.92&      0.95 \\
\textit{w3}&   1189&           17061&       $6.06 \times 10^{9}$&   0.96&      0.99 \\ \hline
\end{tabular}
\begin{flushleft}Note: The daily clickstreams is obtained by summing up the number of unique users over all edges in a clickstream network.
\end{flushleft}
\label{tab.1}
\end{center}
\end{table}

\subsection*{The scaling pattern across language communities}

In the last section, we ignored the differences between users in investigating the scaling property of the clickstream networks. However, this is a naive assumption concerning the different preferences of users in the Web browsing \cite{meiss2010agents,qiu2005analysis}. Among the various demographic and psychological factors that contribute to the preferences \cite{fortunato2006topical,pirolli2007information}, we chose to control the linguistic variance in the further investigation of the scaling property. Specifically, we divided the clickstream networks into language-based website communities and then observed the scaling pattern across the communities.

\begin{figure}[!ht]
\begin{center}
\includegraphics[scale=0.6]{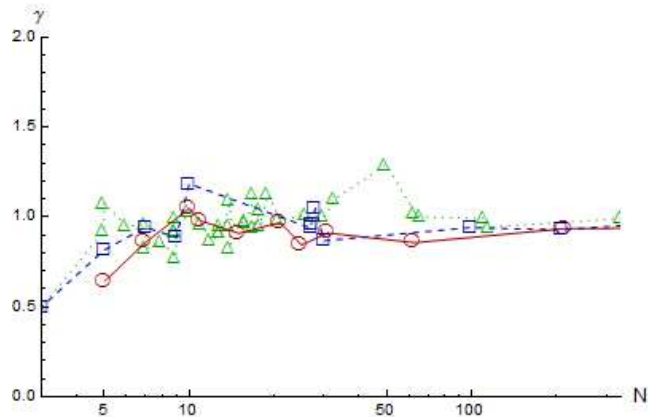}
\end{center}
\caption{{\bf The change of $\gamma$ with community size $N$.} The data points of the three networks are plotted in blue squares ($w1$), red circles ($w2$), and green triangles ($w3$), respectively. The y axis is plotted in the linear scale and the x axis is plotted in the base-$e$ log scale. }
\label{fig.5}
\end{figure}

With the help of the AlchemyAPI (http://www.alchemyapi.com/), which turned out to be very efficient in identifying the languages used by sites, we detected $16$ language communities from $w1$, $17$ from $w2$, and $50$ from $w3$. In Table.\ref{tab.2} we presented the result of $w2$ as an example (the result of the rest two networks were given in the Supplementary Materials). Please note that the communities listed in Table.\ref{tab.2} is less than those given by Figure.\ref{fig.1}, because there were several communities within which the clickstreams were too few to support an effective estimation of $\gamma$. As suggested by Table.\ref{tab.2} and Figure.\ref{fig.5}, in most of the communities there existed the relationship ${C_{i}}\sim{A_{i}}^{\gamma} (\gamma <1)$, and the value of $\gamma $ seem to be invariant of community size. It means that these communities share the common decentralized structure with the entire network. This finding also implies that, despite a variety of demographic and psychological that shapes the preferences of users \cite{fortunato2006topical,pirolli2007information}, there are universal regularities in collective surfing behavior \cite{pirolli2007information}.

We noted that a majority of clickstreams across communities occurred between the English community and non-English communities (Figure.\ref{fig.1}). This is because users generally use no more than two languages (the mother language and English) in surfing the Web. As a consequence, the clickstream network formed a "wheel-like" structure. Whether this structure contributes to the discussed scaling property is an interesting question worth further investigation.

\begin{table}[!ht]
\begin{center}
\caption{The scaling exponent across language communities in $w2$. }
\begin{tabular}{|c|c|c|c|c|c|c|c|c|}
\hline
Community &         $N_{sites}$ & $N_{edges}$  & Daily clickstreams &  $\gamma$& $R^{2}$ of $\gamma$\\\hline
\textit{English}&    516&           6188&       $8.64 \times 10^{9}$&   0.94&      0.94\\
\textit{Chinese}&    214&           2130&       $3.18 \times 10^{9}$&   0.94&      0.77\\
\textit{Japanese}&    63&            481&       $4.83 \times 10^{8}$&   0.86&      0.88\\
\textit{Portuguese}&  31&            115&       $4.48 \times 10^{7}$&   0.91&      0.83\\
\textit{French}&      25&             57&       $1.12 \times 10^{7}$&   0.84&      0.57\\
\textit{Russian}&     21&             94&       $8.50 \times 10^{7}$&   0.97&      0.94\\
\textit{German}&      15&             64&       $1.78 \times 10^{7}$&   0.91&      0.76\\
\textit{Korean}&      11&             53&       $6.11 \times 10^{7}$&   0.98&      0.84\\
\textit{Polish}&      10&             43&       $1.72 \times 10^{7}$&   1.05&      0.91\\
\textit{Vietnamese}&   7&             25&       $8.70 \times 10^{6}$&   0.86&      0.61\\
\textit{Thai}&         3&              6&       $1.67 \times 10^{6}$&   0.31&      0.71\\ \hline
\end{tabular}
\begin{flushleft}The daily clickstreams is derived by summing up the number of unique users over all edges in the clickstream network.
\end{flushleft}
\label{tab.2}
\end{center}
\end{table}


\subsection*{The robustness of the scaling pattern}

The necessity of the work presented in this section is twofold. Firstly, to overcome the limitations of the currently used data sets. As Alexa only provides the top ten inbound and outbound clickstreams for each site, we have to ignore the rest of the potential clickstreams in the data analysis. If the discussed scaling pattern is sensitive to this missing of clickstreams, our conclusion would probably be biased and thus can not be generalized to a larger scale of observation. Secondly, by testing the robustness of the scaling pattern against network reconstructions we may obtain the understanding towards the mechanism leading to the observed pattern.

In the current study, we investigated the robustness of the scaling relationship against two types of network reconstructions, the selective removal of clickstreams and the reshuffling of edges and weights. In both reconstructions we used four statistics to characterize the scaling pattern, including $\gamma$, $R^{2}$, $\rho$, and $D$. $\gamma$ and $R^{2}$ are the best fitted parameter and the explained variance of the OLS regression, respectively. $\rho$ is the Pearson correlation coefficient between $log(C_{i}/A_{i})$ and $log(A_{i})$ \cite{warton2006bivariate}. The reason for introducing $\rho$ is because the scaling exponent $\gamma$ was always close to $1$ in the current study, to focus on the non-linear nature of the data we removed the linear dependence between $A_{i}$ and $C_{i}$ by calculating $C_{i}/A_{i}$ and then observed the direction and strength of the correlation. If $C_{i}$ is irrelevant of $A_{i}$ or has a trivial, linear relationship with $A_{i}$, $\rho$ will approximate $0$; if there exists a relationship ${C_{i}}\sim{A_{i}}^{\gamma} (\gamma < 1)$, which also reads as ${log(C_{i}/A_{i})}\sim(\gamma -1) {log(A_{i})} (\gamma -1 < 0)$, $\rho$ will deviate from $0$ negatively. $D$ is the Kolmogorov-Smirnov statistic that quantifies the distance between two empirical distribution functions \cite{smirnov1948table,clauset2007power}. It can be used to determine whether two data sets come from populations with the same distribution, which is called the KS test \cite{smirnov1948table}. We calculated $D$ between the distributions of the empirical and predicted values of $C_{i}$ (the latter was ${A_{i}}^{\gamma}$, in which $\gamma$ was obtained in the OLS regression) and compared it with $0.035$, the expected value of $D$ corresponding to a confidence level equals $0.1$ (as suggested by \cite{clauset2007power}) and a sample size equals $1200$ \cite{smirnov1948table}. The value of $D$ smaller than $0.035$ was treated as a ``good" result since we could not reject the null hypothesis that the empirical and predicted values of $C_{i}$ were from the same distribution, in other words, the prediction of $C_{i}$ by ${A_{i}}^{\gamma}$ was validated. Please note that we were using the most rigorous criterion of the KS test in setting the sample size as $1200$, because the number of nodes in the three clickstream networks and their reconstructed versions was actually smaller than this number and thus allowed a larger expected value of $D$ \cite{smirnov1948table}. Finally, we would like to stress that among the four statistics, the KS statistic was the only one based on formal statistical tests \cite{smirnov1948table} thus should be treated as the most important criterion.

\begin{figure}[!ht]
\begin{center}
\includegraphics[scale=0.55]{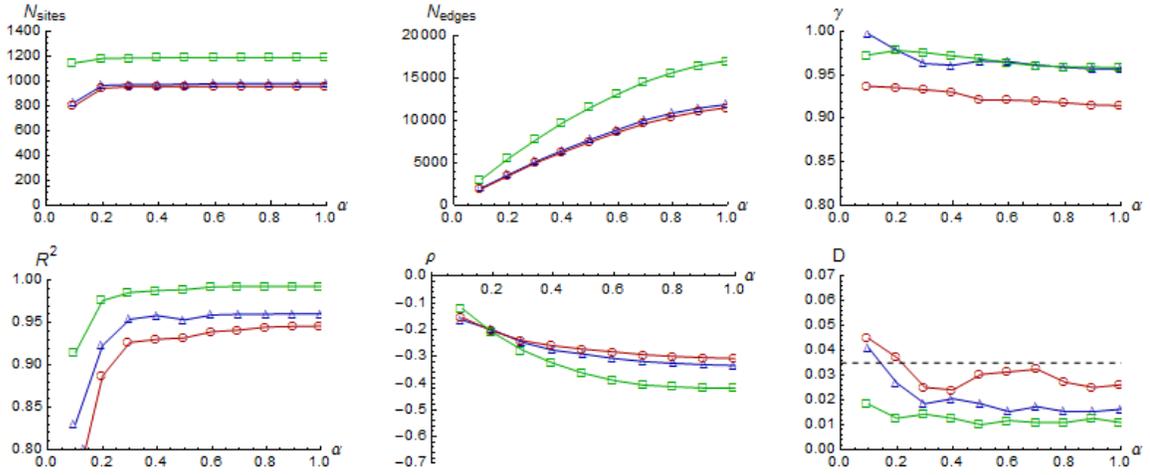}
\end{center}
\caption{{\bf The change of the number of nodes, the number of links, $\gamma$, $R^{2}$, correlation, and KS statistic with the increase of $\alpha$.} The data points from three networks are plotted in different colors: blue squares for $w1$, red circles for $w2$, and green triangles for $w3$. The dashed, black line in the last figure shows the critical value of KS statistics as 0.035 given the condition of 1200 sample size and 0.1 level of confidence.}
\label{fig.6}
\end{figure}

We called the first reconstruction ``backbone network analysis" \cite{foti2011nonparametric}, in which we gradually removed edges of small weights from a network and observed the change of the statistics. Specifically, we defined $0 \leq \alpha < 1$ as the portion of the edges to be kept. For a given $\alpha$, we removed from every node $1-\alpha$ incoming and outgoing edges of the least weights. As shown by Figure.\ref{fig.6}, the scaling pattern was very stable against the removal of edges. In particular, the values of $\gamma$ and $R^{2}$ did not change much while the networks lost as much as $70\%$ edges as $\alpha$ decreased from $1$ to $0.2$. More importantly, during this process the scaling relationship was validated by the KS test. Actually, we can even try to forecast the scaling pattern given the condition that more clickstreams were retrieved: the value of $\gamma$ would probably be smaller and the fitting is likely to be better (to be indicated by the smaller $D$ and the larger $R^{2}$). In sum, our finding of $\gamma<1$ in the clickstream networks is possible to be generalized into the larger scale of observation.

The second reconstruction was called ``reshuffling analysis". In this reconstruction we reshuffled the clickstream networks (in the ways given by Table.\ref{tab.3}) to examine the contributions of the linking structure and weights in forming the scaling relationship.

\begin{table}[!ht]
\begin{center}
\caption{The combinations in the reshuffling. }
\begin{tabular}{|c|c|c|c|}
\hline
                        & Original weights   & Randomly shuffled weights  & Uniformly distributed weights   \\\hline
Original links          &    \textit{w1}/\textit{w2}/\textit{w3}&         \textit{a}&                        \textit{b} \\
Randomly shuffled links &    \textit{c}&                                  \textit{d}&                        \textit{e}  \\
Randomly connected links&    \textit{f}&                                  \textit{g}&                        \textit{h}  \\ \hline
\end{tabular}
\label{tab.3}
\end{center}
\end{table}

Please note that the ``randomly shuffled links" was different from the ``randomly connected links" in Table.\ref{tab.3}, because the former kept the long-tail degree distribution of the original network (as shown in the Supplementary Materials) whereas the latter lead to a binomial degree distribution. In particular, in generating randomly connected links we selected $w$ pairs of numbers (with replacement) from $n$ unique numbers randomly, in which $w$ and $n$ were the number of links and nodes of the network to be reconstructed, respectively. The readers who are familiar with the theories of complex networks would find that we were actually creating Erdos-Renyi random graphs, whose degree distribution is binomial \cite{watts1998collective}. Similarly, the ``randomly shuffled weights" was different from the ``uniformly distributed weights", for we permuted the order of weights and kept their long-tail distribution in the former combination, but created new weights uniformly distributed (between the minimum and maximum values of the original weights) in the latter.

\begin{figure}[!ht]
\begin{center}
\includegraphics[scale=0.5]{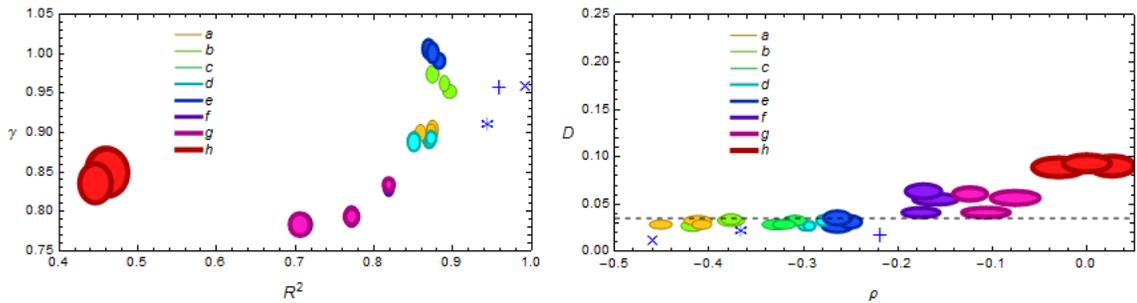}
\end{center}
\caption{{\bf The mean and standard deviation of statistics of interested of the eight combinations in the reshuffling.} The dashed, black line in the left figure shows the critical value of KS statistics as 0.035 given the condition of 1200 sample size and 0.1 level of confidence.}
\label{fig.7}
\end{figure}

For each of the combinations listed in Table.\ref{tab.3}, we ran $100$ times of simulations and recorded the mean and standard deviation of the aforementioned four statistics. After that, we plotted $\gamma$ vs. $R^{2}$ and $\rho$ vs. $D$ as given by Figure.\ref{fig.7}, in which the center of the disks indicated the means and the radius in the corresponding direction reflected the standard deviations. We plotted the results of the combinations in different colors and edge styles and marked the results of the three original networks by ``$+$" ($w1$), ``$*$" ($w2$), and ``$\times$" ($w3$). It turned out that across the three clickstream networks, the original networks always had the smallest $D$ and largest $R^{2}$. In examining the scaling pattern by the KS test, we ruled out three combinations whose values of $D$ were greater than $0.035$, including $f$, $g$, and $h$. From previous discussions on ``randomly connected links" we know that these types of reconstructed networks shared the same binomial degree distribution \cite{watts1998collective}. Therefore, we can naively conclude that the change of the degree distribution from long-tail to binomial blurred the scaling pattern, whereas the change of the weight distribution did not. In other words, the scaling pattern was determined by the topological structure of the clickstream networks.

Compared with $f$, $g$, and $h$, the rest of combinations were close to the original networks in terms of $D$ or other statistics. If we continuously lower the level of confidence in the KS test (e.g., to $0.05$), it is possible to ruled out more combinations. However, we would like to suggest that, before a comprehensive understanding of the mechanism leading to the scaling pattern is achieved, such a tuning is very trivial and does not provide much insight. Therefore, in the current study we would rather stop at the conclusion that topological structure matters in the forming of scaling patterns and leave the contribution of other factors, e.g., the distribution of weights, as a open question.

\section*{Discussion}

We studied collective browsing behavior from a flow network perspective. We defined $C_{i}$ as a measure of the impact of websites $i$ on other sites through users' collective, continuous surfing activities and found it scaled to website traffic $A_{i}$ with an exponent smaller than $1$. This pattern unrevealed the decentralized structure of the three clickstream networks. Further, we found that this scaling pattern appeared universally across language-based communities within the clickstream networks and that the value of $\gamma$ was independent of communities size. Finally, we examined the stability of the scaling pattern against the reconstructions of the clickstream networks. It turned out that the scaling relationship was robust against the selective removal of edges but sensitive to the permutation on the linking structure.

Our finding has relevant theoretical and practical consequences. Although the ``rich-get-richer" paradigm has been widely accepted as a mechanism of hyperlink formations since Barabasi et al. \cite{barabasi1999emergence}, we should not simply assume that this paradigm also suits the dynamics of collective surfing behavior \cite{cho2004impact,introna2000defining}. It is already pointed out in \cite{fortunato2006topical} that the traffic of websites scaled to its number of inbound links with an exponent approaches $0.8$. In this work we found the sublinear relationship between the impact and the traffic. Put these findings together, we can conclude that the survival probability of small sites in the Web ecological system is higher than what was suggested by their in-degree \cite{barabasi1999emergence} or the Page Rank values based on hyperlink structure \cite{page1999pagerank,fortunato2006topical}. Moreover, we would like to emphasize that it is only by studying empirical clickstream networks can the rich interactions between sites be comprehensively understood.

The found scaling relationship provide a quantitative prediction of the impact of a website from its traffic. Online advertising usually measures the impact of websites by their traffic \cite{rosen2004website,tan2007empirical}, but our study offers a more precise calculation of the impact of sites based on their role in the circulation of clickstreams. This approach has potential application in the estimation of the value of sites and also the planning of online marketing campaigns.

\section*{Acknowledgments}


\bibliography{howthewww}

\end{document}